\documentclass[10pt]{iopart}
\usepackage{graphicx}   
\usepackage{latexsym}   
\usepackage{enumerate}
\usepackage{tabularx}
\usepackage{amssymb}
\usepackage[colorlinks=true,linktocpage=true,linkcolor=blue,citecolor=blue,allcolors=blue]{hyperref}

\begin{document}
\title[]{Coalescence, the thermal model and multi-fragmentation:\\ The energy and volume dependence of light nuclei production in heavy ion collisions}

\author{Paula Hillmann$^{1,2,3,4}$, Katharina Käfer$^1$, Jan Steinheimer$^5$, Volodymyr Vovchenko$^6$, Marcus~Bleicher$^{1,2,3,4}$ }

\address{$^1$ Institut f\"ur Theoretische Physik, Goethe Universit\"at Frankfurt, Max-von-Laue-Str. 1, D-60438 Frankfurt am Main, Germany}
\address{$^2$Helmholtz Research Academy Hesse for FAIR (HFHF), Campus Frankfurt,Max-von-Laue-Str.  12, 60438 Frankfurt am Main, Germany}
\address{$^3$ GSI Helmholtzzentrum f\"ur Schwerionenforschung GmbH, Planckstr. 1, 64291 Darmstadt , Germany}
\address{$^4$ John von Neumann-Institut f\"ur Computing, Forschungszentrum J\"ulich,
52425 J\"ulich, Germany}
\address{$^5$ Frankfurt Institute for Advanced Studies, Ruth-Moufang-Str. 1, 60438 Frankfurt am Main, Germany}
\address{$^6$ Nuclear Science Division, Lawrence Berkeley National Laboratory, 1 Cyclotron Road,  Berkeley, CA 94720, USA}

\date{Aug. 25, 2021}
\begin{abstract}
We present results of a phase space coalescence approach within the UrQMD transport and -hybrid model for a very wide range of beam energies from SIS to LHC. The coalescence model is able to qualitatively describe the whole range of experimental data with a fixed set of parameters. Some systematic deviations are observed for very low beam energies where the role of feed down from heavier nuclei and multi-fragmentation becomes relevant. The coalescence results are mostly very close to the thermal model fits. However, both the coalescence approach as well as thermal fits are struggling to simultaneously describe the triton multiplicities measured with the STAR and ALICE experiment. The double ratio of $tp/d^2$, in the coalescence approach, is found to be essentially energy and centrality independent for collisions of heavy nuclei at beam energies of $\mathrm{E_{lab}}> 10 A$ GeV. On the other hand the clear scaling of the $d/p^2$ and $t/p^3$ ratios with the systems volume is broken for peripheral collisions, where a canonical treatment and finite size effects become more important.
\end{abstract}

\submitto{\jpg}

\maketitle

\ioptwocol

\section{Introduction}
The production of light nuclei from a hot and dense system is an interesting topic of study to shed light on both, the early universe evolution (especially the synthesis of light elements) as well as the freeze-out properties of QCD matter in relativistic nuclear collisions~\cite{Andronic:2010qu,Sun:2018jhg,Braun-Munzinger:2018hat}. 
While in the early universe the chemical equilibrium is maintained for a long time and down to temperatures of order of a few MeV, 
the fireball created in heavy ion collisions falls out of chemical and kinetic equilibrium at much higher temperatures. In fact, for the highest beam energies, the last particle scatterings between hadrons occurs at temperatures of around 100 MeV~\cite{Reichert:2020yhx},
which is much higher than the binding energies of nuclei. Still the abundances of these nuclei can be well estimated by assuming that they are created from a system in chemical equilibrium at around 155 MeV. One has to keep in mind that this is drastically different from the standard scenario of nucleosynthesis in the universe, there the typically temperatures are around $0.1-1$~MeV, i.e. of the order or lower than the binding energy of the clusters.
The current explanation for this apparent discrepancy is that, even though the observed light nuclei seem to be  formed after the kinetic freeze out of their constituents, before their emission a multitude of (unobservable) formation and dissociation processes of the clusters happen \cite{Neidig:2021bal}. 
The cluster yield does however not decrease during the expansion from chemical to kinetic freeze-out because due to the rapid nature of the expansion they ''inherit'' the phase space occupancy of their constituents, the protons and neutrons (as e.g. modelled in partial chemical equilibrium approaches \cite{Bebie:1991ij,Xu:2018jff,Vovchenko:2019aoz,Tomasik:2021jfd} or within the kinetic theory description of the hadronic phase~\cite{Oliinychenko:2018ugs,Oliinychenko:2020znl}). In such a scenario, the description of light nuclei production within both the phase space coalescence model as well as the statistical hadronization model would give comparable results. 

On the other hand, the details of the formation of nuclei will leave measurable consequences in the phase space distributions in heavy ion experiments at various beam energies. One example for this would be the clear mass number scaling of the collective flow of light nuclei as predicted by the coalescence model \cite{Yin:2017qhg,Hillmann:2019wlt}. Furthermore, it was suggested that significant variations of the local densities at the point of kinetic freeze out, due to e.g. a phase transition or critical point \cite{Steinheimer:2013gla,Herold:2016uvv}, may also have consequences on the observed nuclei multiplicities. Finally the question occurs how sensitive the formation of nuclei may be on the effects of the volume of the source system and its finite size.  

In this paper we will present results for deuteron and triton production from a phase space coalescence model over a very broad range of beam energies and system sizes and with different underlying dynamical descriptions of the hot and dense phase of the evolution. In particular we will show how robust the deuteron and triton production within this mechanism is when either
a hydrodynamical
description of the bulk evolution is used versus a microscopic transport approach. We can show that the experimentally observed production rates can be very well described by both dynamical approaches and phase space coalescence with only very few parameters, which gives results similar to the predictions of the thermal model. 
In addition, at very low energies, a different model, based on ''hot'' coalescence and subsequent multifragmentation is compared to the standard coalescence and thermal model results. 
In particular we will also discuss the baseline for the double ratio $tp/d^2$ as function of beam energy and as function of system size. In this context we will discuss our results on the system size scaling. 

\section{Model set-up}
To calculate the multiplicities and phase space distributions of deuterons and tritons in high energy nuclear collisions, via a coalescence approach, it is necessary to obtain the phase space distributions of their constituents, the protons and neutrons, at kinetic freeze-out. These can be obtained directly from microscopic model simulations, i.e. a hadronic transport model. 

For our simulations we use the Ultra Relativistic Quantum Molecular transport model (UrQMD) in its default cascade mode and in the hybrid hydro-mode. The UrQMD model in its default setup is based on the binary scatterings of hadrons via a geometric interpretation of the hadronic elastic and inelastic cross sections. The cross sections are taken from experimental data or, if not measured, from model predictions like the one-boson exchange model or the constituent quark model. The scattering processes include elastic reactions, resonance excitations and decays as well as string excitations/fragmentations and strangeness exchange reactions. The model successfully describes experimental data in a wide range of energies and colliding systems. For central collisions at beam energies above $\sqrt{s_{\mathrm{NN}}}> 10$~GeV the collective flow of hadrons is slightly underestimated by the cascade version of the model and a hydrodynamic description of the dense and hot collective phase is necessary \cite{Petersen:2007ca}. For this reason we also use the UrQMD-hybrid model, where the dense phase, after the two incoming nuclei have passed through each other, is simulated with ideal relativistic fluid dynamics, based on the conservation of energy-momentum as well as the net baryon number current. For details of the vanilla UrQMD simulations we refer the reader to \cite{Bass:1998ca,Bleicher:1999xi}.

In the hybrid description, the transition from the fluid description back to the transport description occurs on an iso-energy-density hypersurface  $\epsilon = 3 \epsilon_0$, where $\epsilon_0  \approx 145$~MeV/fm$^3$. The hypersurface is then used to sample hadrons according to the Cooper-Frye equation \cite{Huovinen:2012is} which then continue to interact within the cascade part of the UrQMD model, until reaction cease and kinetic freeze out is reached. For the hydro part we use an equation of state that contains a smooth crossover between a hadron resonance gas and a deconfined quark-gluon-plasma~\cite{Motornenko:2018hjw}.

\begin{table}[t]
\centering
\begin{tabular}{|c|c|c|}
\hline
Probabilities & d & t, $^3$He  \\
\hline
spin-isospin correction & 3/8 & 1/12\\
\hline
\end{tabular}
\begin{tabular}{|c|c|c|c|}
\hline
Parameters & NN & NNN Set I & NNN Set II \\
\hline
$\Delta r_{max}$ [fm] & 3.575 & 4.1 & 4.3\\
\hline
$\Delta p_{max}$  [GeV] & 0.285 & 0.32 & 0.35  \\
\hline
\end{tabular}

\caption{Probabilities and parameters used in the UrQMD  phase-space coalescence. \label{t1}}
\end{table}

\subsection{Production of light nuclei via phase-space coalescence}
Phase-space coalescence is a well established model for the production of light nuclei in heavy ion reactions, which uses the phase space distributions of nucleons, at a time at which all scattering reactions have ceased for the nucleon pair, and folds those distributions with an approximated and correctly normalized wave function of the light nucleus \cite{Schwarzschild:1963zz,Butler:1963pp,Kapusta:1980zz,Bond:1977fd,Nagle:1996vp,Ko:2010zza,Botvina:2014lga,Botvina:2016wko,Sombun:2018yqh,Zhao:2021dka,Sun:2020uoj,Scheibl:1998tk,Glassel:2021rod}. Such an approach would strictly be correct only, if the microscopic description of the nuclear reactions includes a proper relativistic treatment of the nuclear forces, e.g. via a relativistic quantum molecular dynamics approach. Since most descriptions of ultra relativistic heavy ion collisions employ only a classical cascade description of the hadronic freeze out phase, the wave-function coalescence employed here becomes somewhat ambiguous. Nevertheless, the results for cluster production do generally not depend on the details of the wave function (i.e. the results are not sensitive on whether one uses, e.g. for the deuteron, the exact Hulthen wave function, a harmonic oscillator or a box profile \cite{Nagle:1996vp}). Here, we employ a box profile in $\Delta p_{ij}$ and $\Delta r_{ij}$ for the phase space coalescence of light nuclei which is described in \cite{Sombun:2018yqh} and was shown to yield a successful description of deuterons at various energies. 

Here, the coalescence probability of nucleons is calculated at their last space-time points of interaction, the kinetic freeze-out. In the case of the deuteron for example, a proton-neutron pair, after their last scattering, is considered a deuteron candidate if the relative momenta and distance in their two-particle center-of-mass frame is smaller than $\Delta r_{max,nn}$ and $\Delta p_{max,nn}$. The probability to form a deuteron within this phase-space volume is then given by the proper statistical spin-isospin projection correction factor \cite{Mattiello:1996gq} which takes into account that only a fraction of possible in-going spin-isospin combinations can actually be realized as a deuteron. The absolute normalization of the number of nuclei created is therefore controlled by the phase-space volume which leaves this approach with an additional degree of freedom. This freedom of the choice of the phase-space volume can be understood as a result of the missing implementation of the actual nuclear forces which would naturally bind the nucleons into nuclei. In other words, since the cascade type transport models can only describe potential (or proto-) nuclear states an additional uncertainty in the phase-space volume for the proto-nucleus arises. However, this uncertainty of the absolute volume can easily be resolved by fitting the parameters $\Delta r_{max,nn}$ and $\Delta p_{max,nn}$ to data at any specific collision system and collision energy. Once the parameters are fixed, the results for all other collision systems at various beam energies can be regarded as parameter free predictions.

The phase-space coalescence approach  described  above was already successfully used to describe deuteron and anti-deuteron production~\cite{Sombun:2018yqh}. This approach has been extended to the production of tritons and helium-3.

\begin{figure}[t]	
\includegraphics[width=0.5\textwidth]{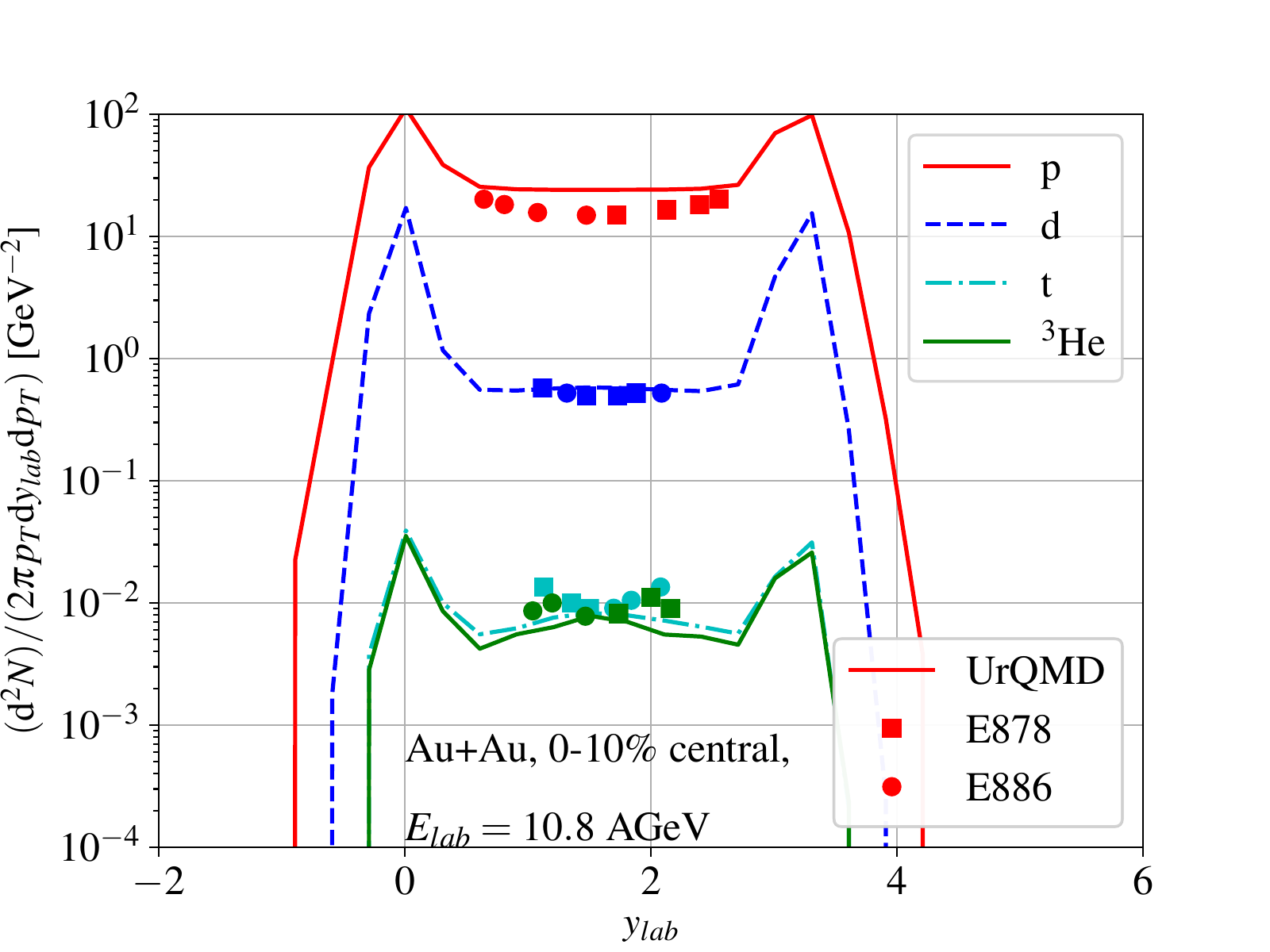}
\caption{[Color online] Invariant yields of protons and light nuclei as function of rapidity for $p_T/A<0.1$ GeV in 0-8\% central Au+Au collisions at $E_{lab}=10.8$ AGeV. The lines denote the UrQMD simulations and the symbols the experimental data~\cite{Bennett:1998be}. Parameter $Set I$ is used.}\label{f2}
\end{figure}

The detailed procedure is as follows, where we assume that a triton or helium-3 is formed by coalescence of a (pre-)deuteron with another nucleon:
\begin{enumerate}
\item As first step we look in the two-particle-rest-frame of each possible two-nucleon pair \footnote{In fact only nucleon pairs (and triplets) with the correct isospin combination, i.e. \textit{pn} for the deuteron and \textit{pnn} for the triton, are allowed.}. If their relative distance $\Delta r=\left|\vec{r}_{n_1}-\vec{r}_{n_2}\right|<\Delta r_{max,nn}=3.575$ fm and momentum distance $\Delta p=|\vec{p}_{n_1}-\vec{p}_{n_2}|<\Delta p_{max,nn}=0.285$ GeV, a two nucleon state is potentially formed with the combined momenta $\vec{p}_{nn}=\vec{p}_{n_1}+\vec{p}_{n_2}$ at position ${\vec{r}_{nn}=(\vec{r}_{n_1}+\vec{r}_{n_2})/2}$. The parameters $\Delta p_{max,nn}$ and $\Delta r_{max,nn}$ correspond to those obtained for the deuteron in \cite{Sombun:2018yqh}.
\item As second step we boost into the local rest-frame of this two nucleon state and any other possible third nucleon. If the conditions of their relative distance $\Delta r=|\vec{r}_{nn}-\vec{r}_{n_3}|<\Delta r_{max,nnn}$ and momentum distance $\Delta p=|\vec{p}_{nn}-\vec{p}_{n_3}|<\Delta p_{max,nnn}$ are fulfilled, a triton (charge equals 1) or helium-3 (charge equals 2) is formed with the probability of $(1/12)\cdot (1/3!)$, the first factor represents the spin-isospin-coupling, and the second is due to different combinations leading to the same $nnn$-state. The momentum of the three nucleon state is then $\vec{p}_{nnn}=\vec{p}_{nn}+\vec{p}_{n_3}$ and the position is $\vec{r}_{nnn}=\frac{1}{3} \left( \vec{r}_{n_1}+\vec{r}_{n_2}+\vec{r}_{n_3} \right)$. 
\item If no third particle is found and the charge equals 1, a deuteron is formed with the probability of $(3/8)\cdot (1/2!)$, the first factor represents the spin-isospin-coupling, and the second is due to different combinations leading to the same nn-state. (For the deuterons, this treatment is the same as in \cite{Sombun:2018yqh}, the extra factor $(1/2!)$ just removes combinatorial double counting in the numerical procedure.)
\end{enumerate}

\begin{figure}[t]	
\includegraphics[width=0.5\textwidth]{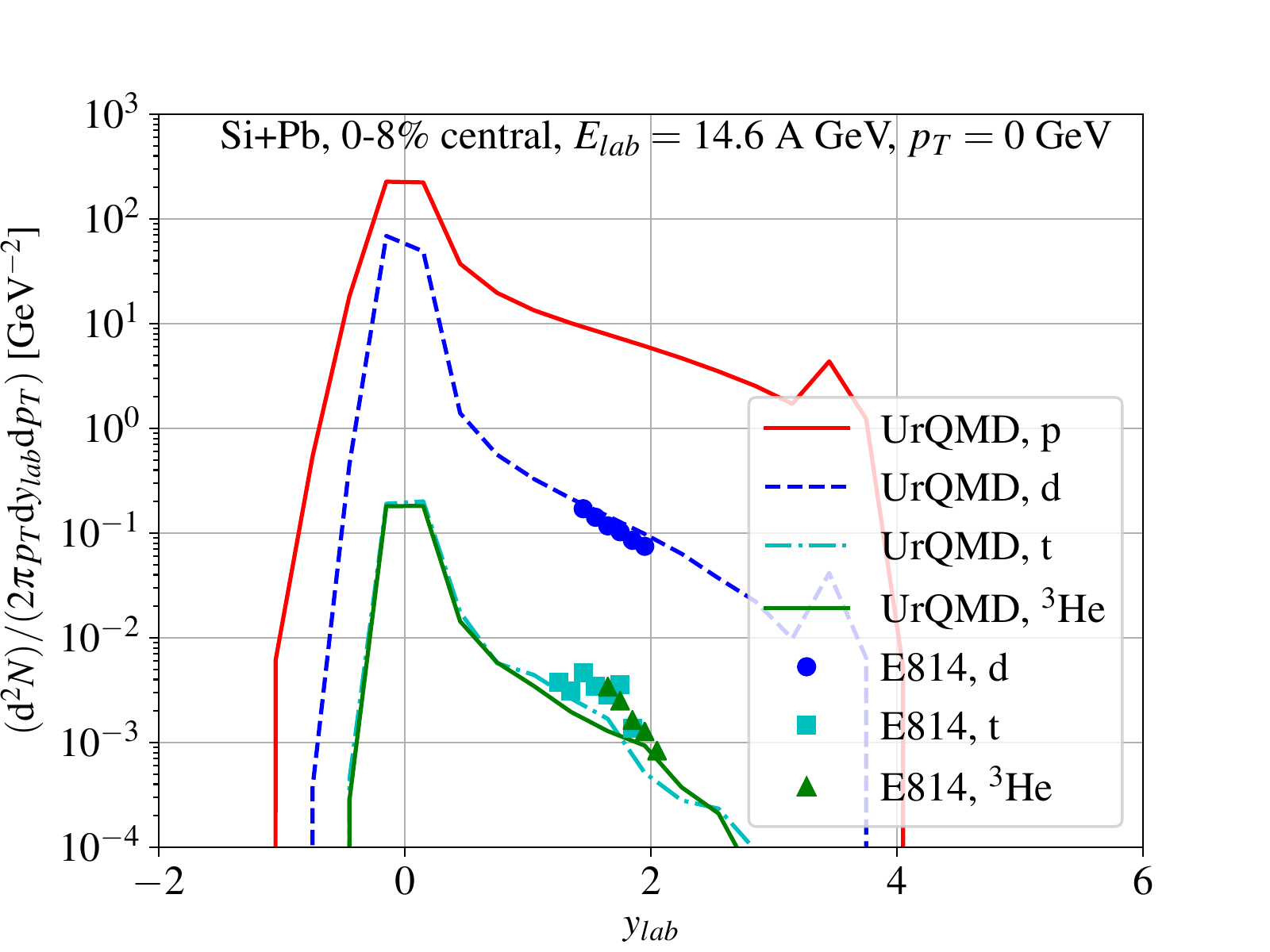}
\caption{[Color online] Invariant yields of protons and light nuclei as function of rapidity for $p_T/A<0.1$ GeV in 0-8\% central Si+Pb collisions at $E_{lab}=14.6$ AGeV. The lines denote the UrQMD simulations and the symbols the experimental data~\cite{Barrette:1994tw}. Parameter $Set I$ is used.}\label{f4}
\end{figure}

\begin{figure}[t]	
\includegraphics[width=0.5\textwidth]{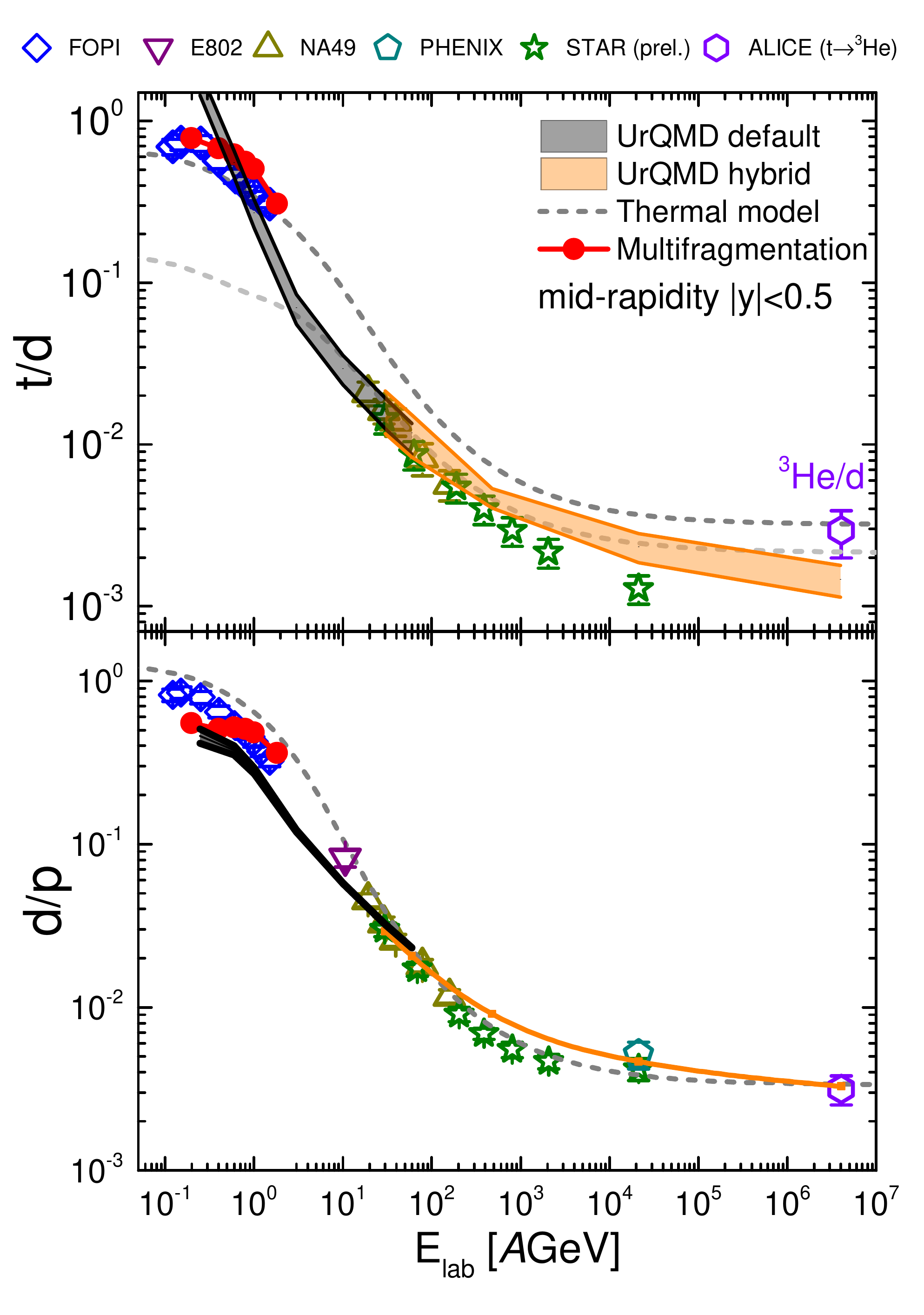}
\caption{[Color online] Ratio of tritons to deuterons (upper panel) and deuterons to protons (lower panel) in mid-rapidity acceptance ($|y|<0.5$) for most central heavy ion collisions at various different beam energies. We show default UrQMD result (black band) with UrQMD hybrid model results using a crossover EoS (orange band). The band depict the uncertainty from using either parameter Set I or II (see text). The UrQMD results are compared to thermal model fits (dashed lines) and Multifragmentation results (red line with circles). Data from different experiments is shown as open symbols. }\label{rat_set1}
\end{figure}

This procedure leaves the two coalescence parameters $\Delta p_{max,nnn}$ and $\Delta r_{max,nnn}$ to be determined. For simplicity, we assume that these parameters are the same for  triton and $^3\mathrm{He}$ and their relative abundances only depend on the different spin-isospin factors as shown in table \ref{t1}. Effects like the different binding energy due to the isospin dependent nuclear forces are neglected.
One should note that for systems where the triton multiplicity is not negligible as compared to the deuteron multiplicity, the order in which the light nuclei are produced in the coalescence procedure becomes important. In our approach we first check for all possible 3 nucleon configurations before deuterons can be produced, as we do not allow the same nucleons to be part of more than one cluster. This means that the more tritons are produced, less p-n pairs will be available for deuteron production.

\begin{figure*}[t]	%
\center
\includegraphics[width=1.0\textwidth]{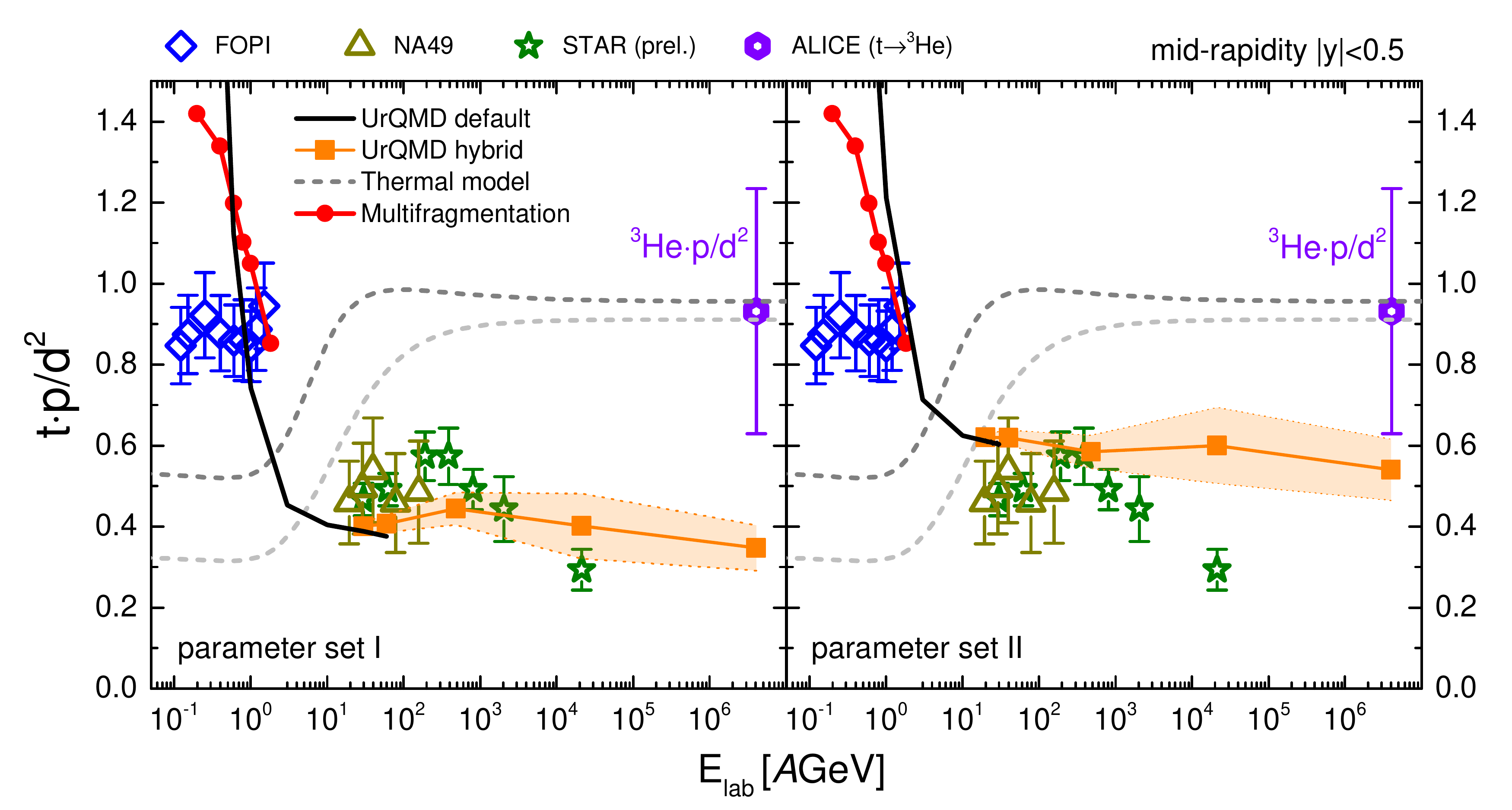}
\caption{[Color online] Double ratio of tritons and protons over  deuterons in mid-rapidity acceptance ($|y|<0.5$) for most central heavy ion collisions at various different beam energies. We show results for both parameters sets I and II. The default UrQMD result are depicted as black line with the UrQMD hybrid model results as orange line with squares. The UrQMD results are also compared to thermal model fits (dashed lines) and Multifragmentation results (red line with circles). Data from different experiments is shown as open symbols. }\label{drat_set1}
\end{figure*}

To fix the coalescence parameters we use the high precision results on the production cross section of triton and $^3\mathrm{He}$ in Au+Au collisions at $E_{lab}=10.8 A$ GeV, as measured by the E878 and E886 collaborations and in central Si+Pb collisions presented by the E814 collaboration. The resulting rapidity distributions using the UrQMD model in cascade mode and the above described procedure using the parameters shown in table \ref{t1}, named \textit{Set I}, are shown in figures \ref{f2} and \ref{f4}. To accommodate for a certain degree of uncertainty in the production cross section of the $A=3$ nuclei and also to show the robustness of the approach to small changes in the coalescence parameters, table \ref{t1} also shows a second set of parameters, dubbed \textit{Set II}. In the following all results presented are using either \textit{Set I} or \textit{II}.

\subsection{Production of light nuclei via hot coalescence and multi-fragmentation}

Usually, as described above, coalescence is done in the freeze-out stage of the reaction by projecting individual nucleon pairs or triples/quadruples on a cluster ground state. However, this is not the only possibility to realize the coalescence process. In \cite{Botvina:2020yfw} a different approach was suggested, namely ''hot coalescence'' followed by multifragmentation. Such an approach is especially important at low beam energies where also intermediate mass clusters are produced. The idea behind ''hot coalescence'' is that one coalesces the nucleons to larger fragments and not directly to the ground states. These intermediate mass fragments still have thermal excitation energy (this is why we call it ''hot coalescence'') and decay via a multifragmentation process producing the observable light clusters, like deuterons, tritons and helium.
 These hot clusters are identified at a given time after the collisions have ceased. A cluster with mass number $A$ is defined as consisting of all nucleons inside a sphere with radius $R = R_0 A^{1/3}$, with $R_0=2.0$ fm, around the center of mass of that cluster. Once identified, the total baryon number and excitation energy of that nuclear cluster is calculated and its decay into stable nuclei is modeled using a microscopic multifragmentation approach. Such a procedure only makes sense if the nucleon density after kinetic freeze out is still large enough so that clusters with large baryon number can be formed, which is only true for the lowest beam energies. Here the number of high mass nuclei produced is significant enough so that the 'hot coalescence' approach makes sense and leads to results that are different to the simple coalescence which ignores the possibility of producing nuclei with large mass number $A>3$.
The detailed implementation and discussion of this process is provided in \cite{Botvina:2020yfw}. 

\subsection{Production of light nuclei in the thermal model}

In the thermal model approach the light clusters are usually incorporated into the partition function as explicit degrees of freedom and their abundances are given by the Boltzmann distribution function, depending only on the temperature and baryo-chemical potential~\cite{Andronic:2010qu,Donigus:2020ctf}.
Under the assumption that the temperatures and chemical potentials for the light clusters are the same as for the nucleons, the thermal model gives an essentially parameter-free prediction of the cluster abundances for a given system, which in many cases is in a very good agreement with the experiment~\cite{ALICE:2015wav,NA49:2016qvu}.
One can also show that certain ratios of primordial yields, like $tp/d^2$, become essentially independent of the temperature and baryochemical potential and reflect only the spin- and isospin-degeneracies.
For the final yields, however, the situation is somewhat more involved since one has to take into account feed-down contributions from decays of baryon resonances~\cite{Oliinychenko:2020ply} and excited nuclei~\cite{Shuryak:2019ikv}.

Here we will confront our coalescence results with the thermal model baseline, which we take from Ref.~\cite{Vovchenko:2020dmv}.
The calculations, performed within the \texttt{Thermal-FIST} package~\cite{Vovchenko:2019pjl}, utilize the chemical freeze-out curve from Ref.~\cite{Vovchenko:2015idt} to fix the temperature and chemical potential at given collision energy. We optionally incorporate the feed-down contributions from known excited states of nuclei, which mainly has the effect of enhancing the yields of tritons and $^3$He relative to deuterons and nucleons.

\section{Results}

First, the ratios of triton to deuteron and deuteron to proton for the mid-rapidity region of central Au+Au reaction is shown in figure \ref{rat_set1} over a very broad range of beam energies. We confront different models with the available experimental data \cite{ALICE:2015wav,PHENIX:2003iij,PHENIX:2004vqi,Zhang:2020ewj,STAR:2019sjh,NA49:2016qvu,E802:1999hit,E877:1999qpr,E864:2000auv,FOPI:2010xrt}. For the highest beam energy, at the LHC, no published data for the triton multiplicity is available from the ALICE experiment yet. However, the $^3$He multiplicity has been published and we will use it as a proxy for the triton. This is reasonable as neither the coalescence model nor the thermal model predict any meaningful difference between triton and $^3$He at this high beam energy. This is also supported by triton from p+Pb collisions measured by ALICE \cite{ALICE:2019fee}. Eventually, our predictions will have to be compared to measurements of the triton once they become available from the ALICE collaboration.

The black solid line corresponds to the cascade version of the UrQMD model plus coalescence and the orange line depicts the UrQMD hydro-hybrid model plus coalescence. The deuteron and triton production is estimated with parameters sets $I$ and $II$ which then result in the grey and orange bands in the figure. The light and dark grey dashed lines are the results for the thermal model fits, where in the dark grey line the feed-down from excited nuclei to the triton and deuteron yields is taken into account. The red line with squares depicts the results from the 'hot coalescence' calculation presented in \cite{Botvina:2020yfw}. 
The various model simulations are compared with experimentally measured ratios from different experiments, shown as open symbols.

One can make several interesting observations: 
\begin{enumerate}
    \item The UrQMD cascade and hybrid model give essentially identical results in the beam energy region between $2 < E_{\mathrm{lab}}< 40 A$ GeV \footnote{We have checked that both models give the same results down to beam energies of $ E_{\mathrm{lab}}=2 A$ GeV. The low beam energy results are simply not shown for better readability of the figures.}. This is a clear indication, that the final ratios mainly depend on the local density at kinetic freeze-out. 
    \item For beam energies above $E_{\mathrm{lab}}>10 A$ GeV, the UrQMD coalescence approach and the thermal model (without feed down) give very similar results for both ratios, where the largest differences occur for the highest (LHC) energy as well as at low beam energies.
    \item All models shown here do not describe well
    the deuteron to proton ratio measured by the FOPI Collaboration at the lowest beam energies. While it can be argued that these energies might be outside the application range of both the coalescence and thermal model,
    it also observed that the multifragmentation approach, which seems to be the most reasonable approach for such low beam energies, also has problems describing the FOPI data.
    \item There appears to be a tension between the triton to deuteron ratio observed by the STAR experiment and the preliminary ratio presented by the ALICE experiment. No model is able to simultaneously describe these two sets of data.\footnote{Differences in the treatment of feed-down corrections have been suggested as a possible source of this difference~\cite{Shao:2020lbq,Oliinychenko:2020znl}. For a further discussion, we refer to \cite{Vovchenko:2020dmv} and references therein.}
\end{enumerate}

As a general trend, at the lowest beam energies, we observe that there is a clear anti-correlation between triton and deuteron production in the coalescence model. As the triton multiplicity is overestimated, due to the lack of higher mass number nuclei which will become important for the lowest beam energies, the deuteron number is suppressed. This effect is less obvious in the 'hot coalescence' approach which included the fragmentation into larger clusters and therefore provided an improved description of the data. Interestingly, the thermal model, including the feed down of excited high mass nuclei, works best for the lowest beam energies while a no-feed-down scenario is preferred for beam energies above $\mathrm{E_{lab}}>10 A$ GeV.

Since the $d/p$ and $t/d$ ratios depend strongly on the net baryon density at freeze out and therefore the baryo-chemical potential of the system, the ratios vary over several orders of magnitude as function of beam energy, which makes a proper comparison and interpretation (apart from the strong $\mu_B$ dependence) difficult.

\begin{figure}
\centering
\includegraphics[width=0.5\textwidth]{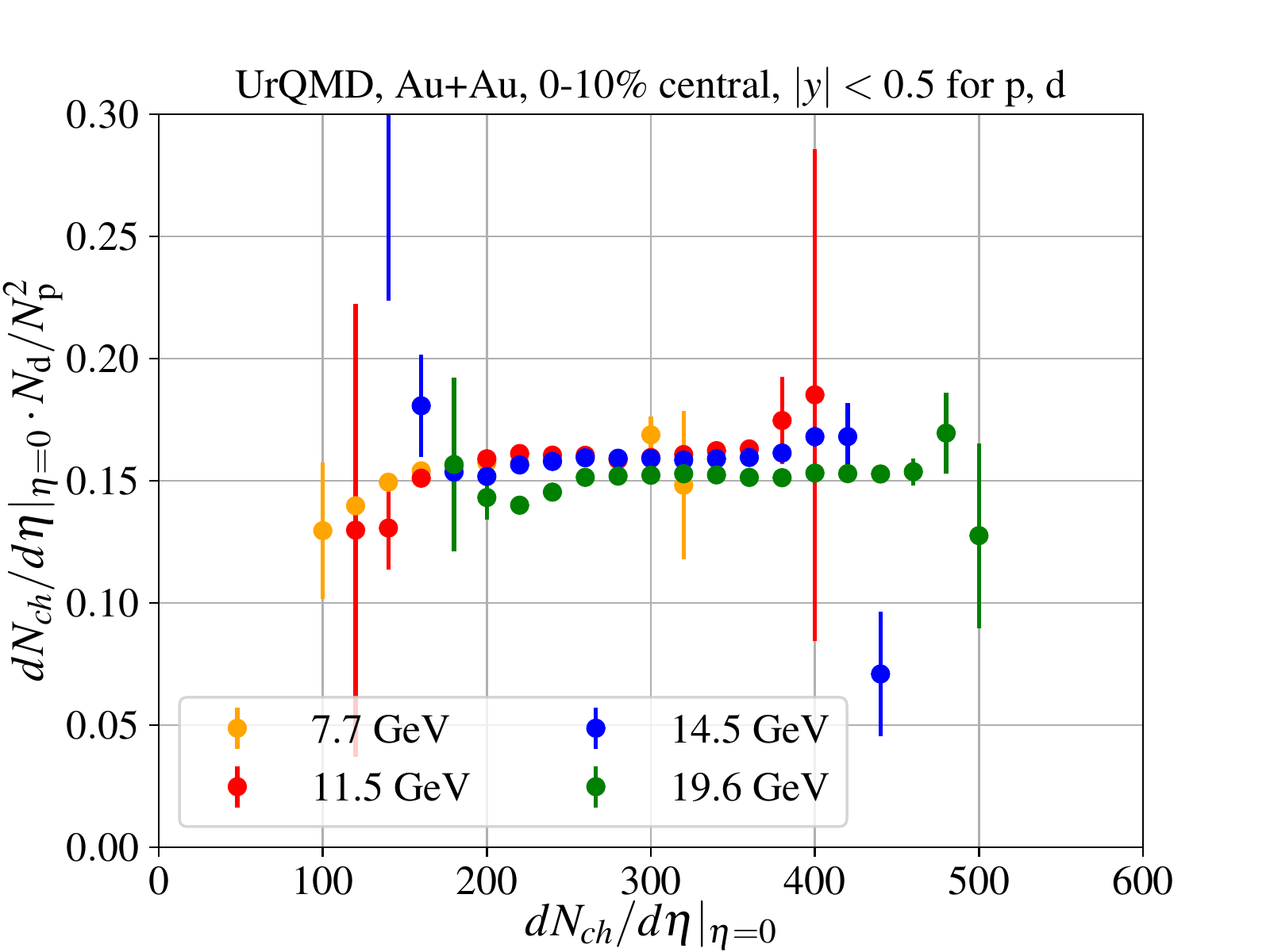}
\caption{Ratio of the deuteron number divided by the squared proton number $d/p^2$, for $|y|<0.5$, scaled with the charged particle multiplicity $dN_{ch}/d\eta |_{ \eta  =0.0}$ as function of the charged particle multiplicity for 0-10\% central Au+Au reactions  at different beam energies as depicted in the Figure.  The UrQMD model in cascade mode is used.}
\label{fig:dpp_volume}
\end{figure}

To remove the trivial $\mu_B$ dependence one can choose a ratio which cancels the explicit dependence on the chemical potential, for example the double ratio $t \cdot p / d^2$. Here $t$, $p$ and $d$ stand for the event averaged multiplicities of the triton, proton and deuterons at mid-rapidity, respectively. This ratio has also been suggested to be sensitive to the neutron number fluctuations in heavy ion collisions \cite{Sun:2018jhg}. However, in terms of a coalescence picture this double ratio can directly be related to the ratio of the so-called two- and three-particle coalescence factors $B_2$ and $B_3$ by 

\begin{equation}
    \frac{tp}{d^2} \propto \frac{B_3}{B_2^2}
\end{equation}

The two- and three-particle coalescence factors are given by:
\begin{eqnarray}
  B_2 &=&  \frac{ \int dp^3 \left. \frac{d^3N_{\mathrm{d}}}{dp^3}\right| _{\vec{p}_{\mathrm{d}}=2\vec{p}_{\mathrm{p}}}}{\left( \int dp^3 \left. \frac{d^3N_{\mathrm{p}}}{dp^3}\right| _{\vec{p}_p}\right) ^2}\, , \\
  B_3 &=& \frac{\int dp^3 \left. \frac{d^3N_{\mathrm{t}}}{dp^3}\right| _{\vec{p}_{\mathrm{t}}=3\vec{p}_{\mathrm{p}}}}{\left( \int dp^3 \left. \frac{d^3N_{\mathrm{p}}}{dp^3}\right| _{\vec{p}_p}\right) ^3}\, .
\end{eqnarray}
One should note that theoretically $B_2$ and $B_3$ can also be defined without integrating over the momenta,
i.e. they can relate the cluster yield at one momentum per nucleon to the nucleon yield at the same momentum. 
However, most of the time one integrates the particle yields over a range in transverse momenta and/or a certain rapidity interval and obtains an integrated $B_2$ and $B_3$. 
Here, we consider these factors, integrated over one rapidity unit $|y|<0.5$ and full $p_T$.

\begin{figure}
\centering
\includegraphics[width=0.5\textwidth]{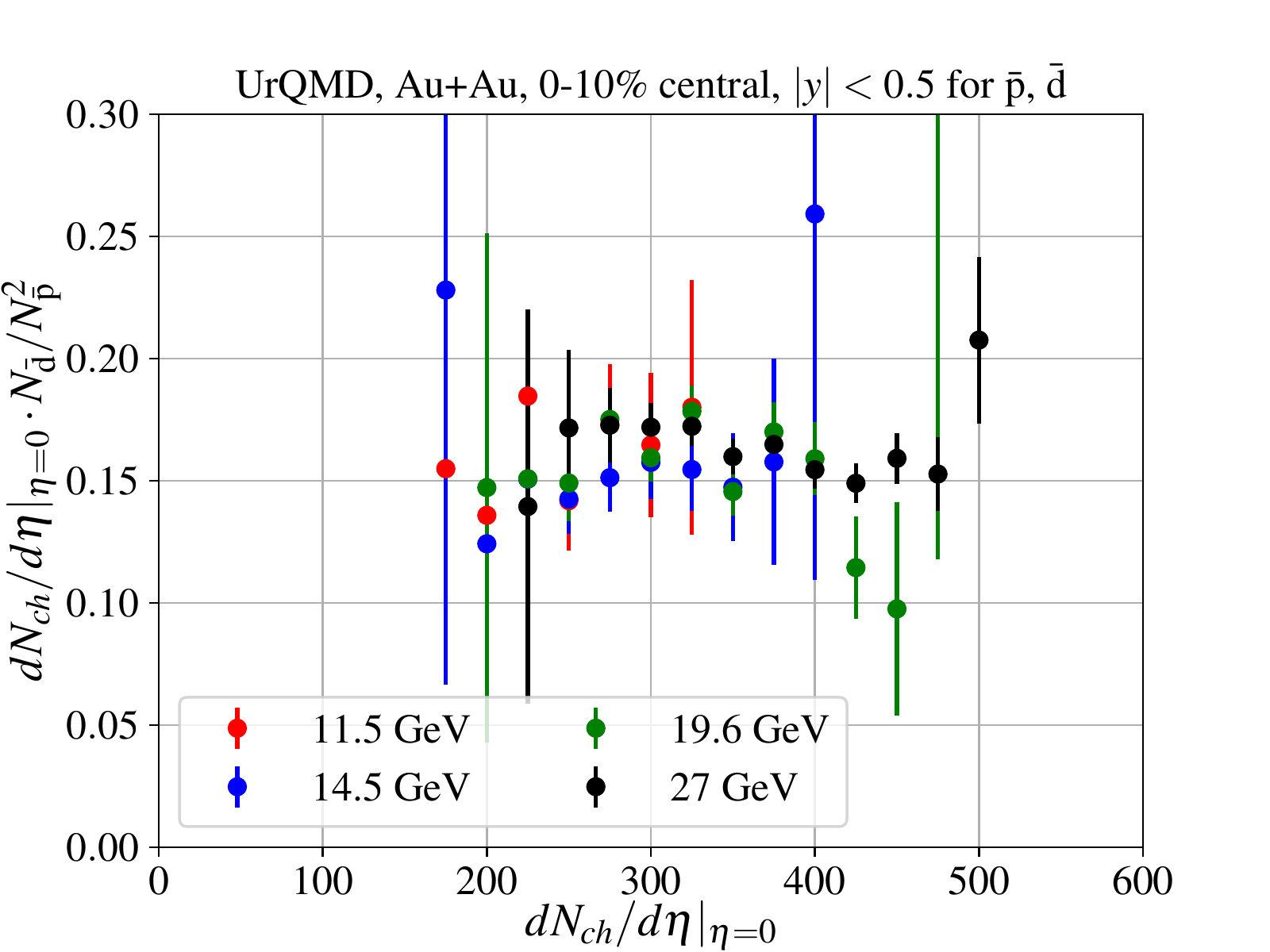}
\caption{Ratio of the anti-deuteron number divided by the squared anti-proton number $d/p^2$, for $|y|<0.5$, scaled with the charged particle multiplicity $dN_{ch}/d\eta |_{ \eta =0.0}$ as function of the charged particle multiplicity for 0-10\% central Au+Au reactions at different beam energies as depicted in the Figure. The UrQMD model in cascade mode is used.}
\label{fig:anti_dpp_volume}
\end{figure}

The results for this double ratio for the different models under investigation is shown in figure \ref{drat_set1} (left: set I, right: set II of triton coalescence parameters). The black solid line corresponds to the cascade version of the UrQMD model plus coalescence and the orange line depicts the UrQMD hydro-hybrid model plus coalescence. The light and dark grey dashed lines are the results for the thermal model fits, where in the dark grey line the feed-down of larger nuclei to the triton yield is taken into account. The red line with squares depicts the results from the multifragmentation calculation presented in  \cite{Botvina:2020yfw}. The data are shown as open symbols. For both parameter sets the double ratio, in the coalescence approach, changes only mildly as function of the beam energy and is essentially constant. Only for very low beam energies a sudden increase is observed which is due to the interplay between triton and deuteron production. As described above, the triton multiplicity becomes as large as the deuterons and since the total baryon number is conserved, a given triton and $^3He$ multiplicity only leaves a reduced number of p-n pairs available for deuteron production.
The thermal model results, on the other hand, are mainly sensitive to the change in temperature along the freeze out curve, which regulates the proton feed-down from baryon resonances.
The temperature increases and then saturates towards higher beam energies, and this mirrors the behavior of the double ratio in the thermal model.
The multifragmentation approach shows a very strong energy dependence and generally the largest achieved values of the double ratio, similar to that of the coalescence model. The main reason for this behaviour is again the reduction of the deuteron yield in this approach due to the increased production of larger nuclei, which gets amplified due to the square in the double ratio.

Both, the normal and 'hot' coalescence overestimate the double ratio, measured by FOPI, for the lowest beam energies $\mathrm{E_{lab}}<1 A$ GeV. This indicates that subtleties in the multifragmentation process which influence the relative multiplicities of the different nuclei, are very important and not fully captures by either model. 

None of the presented models is able to describe all of the available measurements. Again, a significant tension is observed between the highest energy STAR and ALICE data. The low energy FOPI data seem to favor a multifragmentation scenario, though also here strong deviations from the model are apparent.

Within the presented UrQMD and hybrid approach it will be possible to study the effects of explicitly including fluctuations due to a phase transition, which will be done in a future study. The current results can be understood as a baseline in the case of no explicit correlations between the baryons.

\begin{figure}[t]
    \centering
    \vspace*{-0.5cm}
    \includegraphics[width=0.5\textwidth]{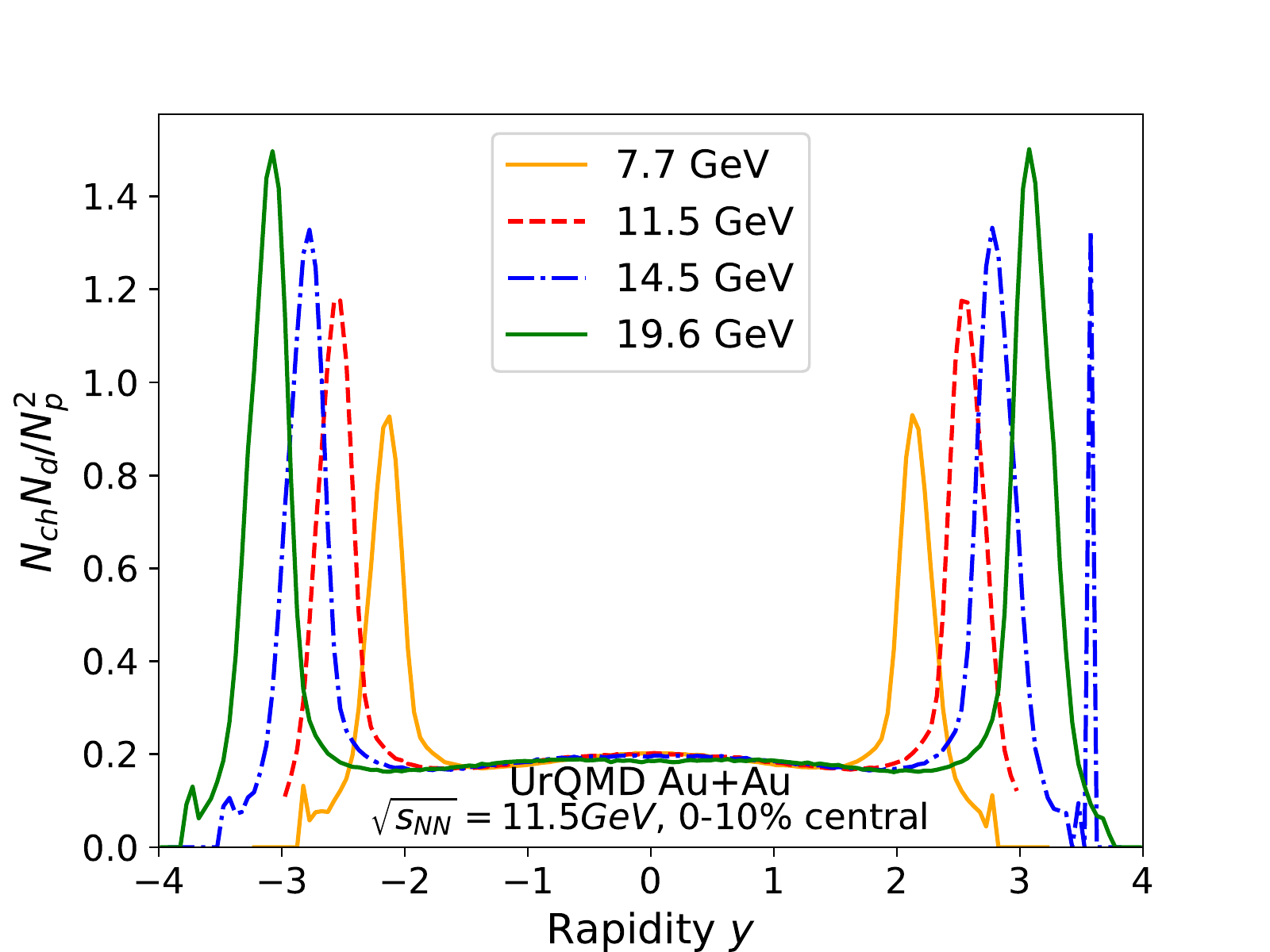}
    \caption{Ratio of deuterons to protons squared multiplied by the number of charged particles as function of the rapidity for central Au+Au reactions at various beam energies. UrQMD in cascade mode is used.}
    \label{fig:dppch_rap}
\end{figure}

\subsection{System size dependence and canonical effects}
It was suggested that the strong decrease of the cluster production (or ratios of light nuclei yields) with energy is mainly driven by the increase in the volume of the source with increasing energy. It was speculated that such a volume increase manifests itself in a scaling of the cluster ratios with the freeze-out volume \cite{Sorge:1995vm} as defined by the charged particle or pion multiplicity. 

In a recent work \cite{Zhao:2021dka} the multiplicity/volume dependence of light nuclei production was studied in a hydro-hybrid approach, similar to the UrQMD model used in this study. There, a possible breaking of the volume scaling, of the coalescence parameters $B_2$ and $B_3$ in peripheral collisions, was discussed.

In the most naive interpretation of the coalescence model, certain ratios like $d/p^2$, where $d$ is the number of deuterons and $p$ is the number of protons in a measured volume, will scale inversely with the volume of the system and thus with the total number of charged particles in said volume (see also \cite{Mrowczynski:1993cx} for an early discussion of this effect and its update in \cite{Kittiratpattana:2020daw}). This volume scaling for deuterons has already been observed in \cite{Sorge:1995vm} for SPS energies (for S+S, S+W and central Pb+Pb reactions). Here we explore, if this scaling also holds for central Au+Au reactions at lower RHIC-BES energies and for different multiplicity bins at fixed centrality. This allows us to pin down, if the deuteron production is indeed driven by the volume effect, when keeping all other parameters essentially fixed. 
At first, we want to study the scaling of $d/p^2$ and (their anti-particles) with the charged particle yield at midrapidity $dN_{ch}/d\eta$ for a fixed beam energy and central collisions. Figure \ref{fig:dpp_volume} shows the $d/p^2$-ratio, at midrapidity, $|y|<0.5$, scaled with $dN_{ch}/d\eta$ for central Au+Au reactions for various beam energies (as entitled in the legend). The same ratio for the anti-particles is shown in Fig. \ref{fig:anti_dpp_volume}.

One observes that the multiplicity scaled $d/p^2$-ratio is essentially energy independent and also multiplicity independent. In addition one observes that the scaled ratio is similar for particle and anti-particles. This suggests two interpretations: I) The deuteron formation process is indeed driven by the volume in this energy range and II) the kinetic freeze-out volume is similar for baryons and anti-baryons, which indicates that the kinetic freeze-out (at these energies) is driven by the pion wind, i.e. scattering reactions involving pions are dominant, and not by baryon-anti-baryon annihilation as suggested in \cite{Mrowczynski:1993cx} for lower energies.

Fig. \ref{fig:dppch_rap} shows the ratio of $d/p^2$ times $N_{ch}$ as function of rapidity for central Au+Au reactions at various beam energies (as can be seen in the legend). The scaling itself is present over an extended range in the central rapidity region, suggesting that cluster formation proceeds at a the same nucleon density over the central region. Towards the fragmentation region, deuteron production increases drastically due to the spectator matter which, of course, does not show the volume scaling. 

Thus, we conclude that volume/multiplicity scaling can be expected for a wide multiplicity range of central collisions and different energies.

\begin{figure}[t!]	
\includegraphics[width=0.5\textwidth]{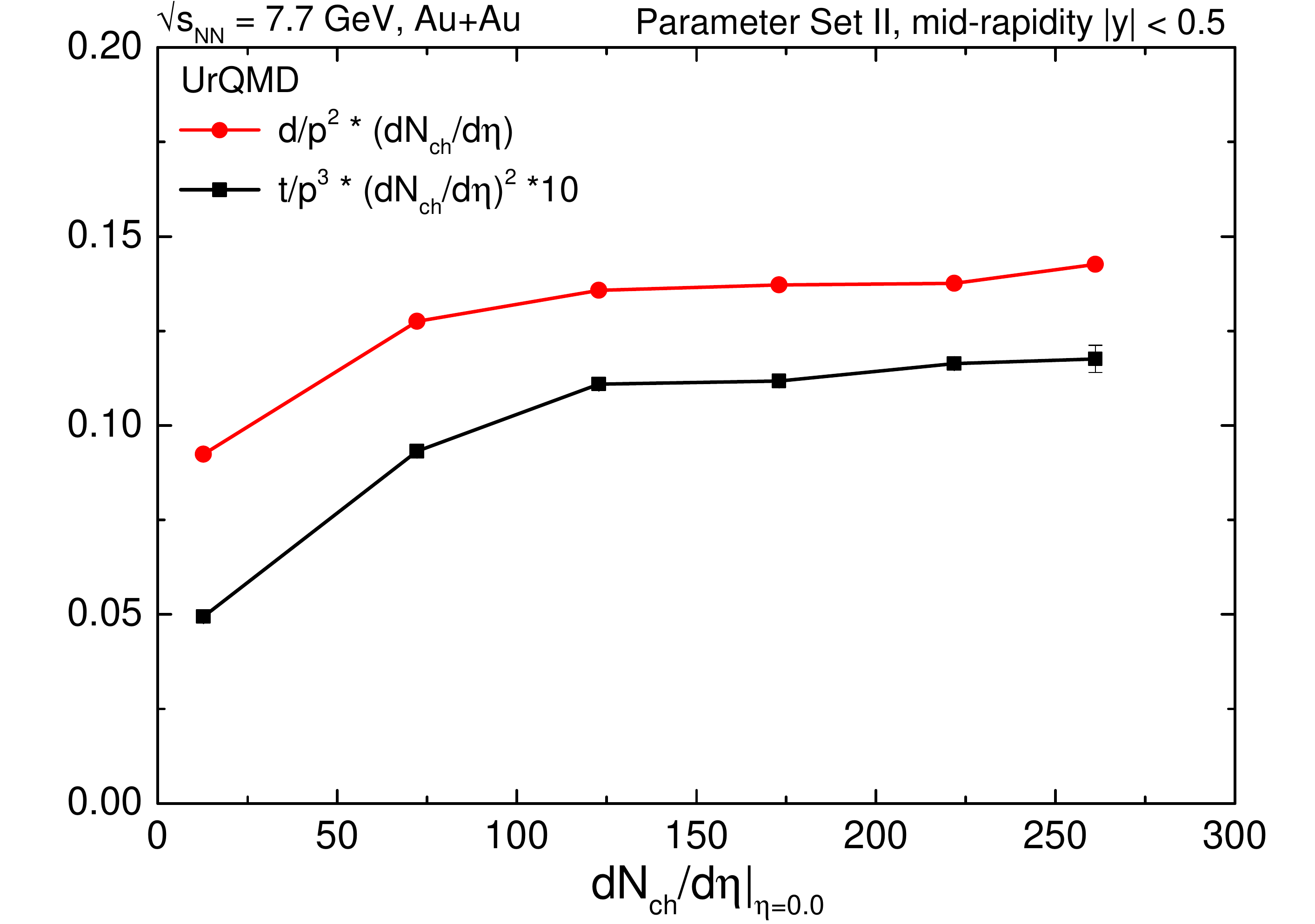}
\caption{[Color online] Centrality dependence of the two ratios d/p$^2$ and t/p$^3$ corresponding to the coalescence parameters $B_2$ and $B_3$ for the mid rapidity region of Au+Au collisions at $\sqrt{s_{\mathrm{NN}}}=7.7$ GeV. Both ratios are scaled with the appropriate power of the net charged multiplicity to remove the expected volume scaling. While central collisions show a clear volume scaling of the coalescence parameters, this scaling is violated for peripheral collisions, where effects of the finite size of the system become important. } \label{cent_sc}
\end{figure}

\begin{figure}[t!]	
\includegraphics[width=0.5\textwidth]{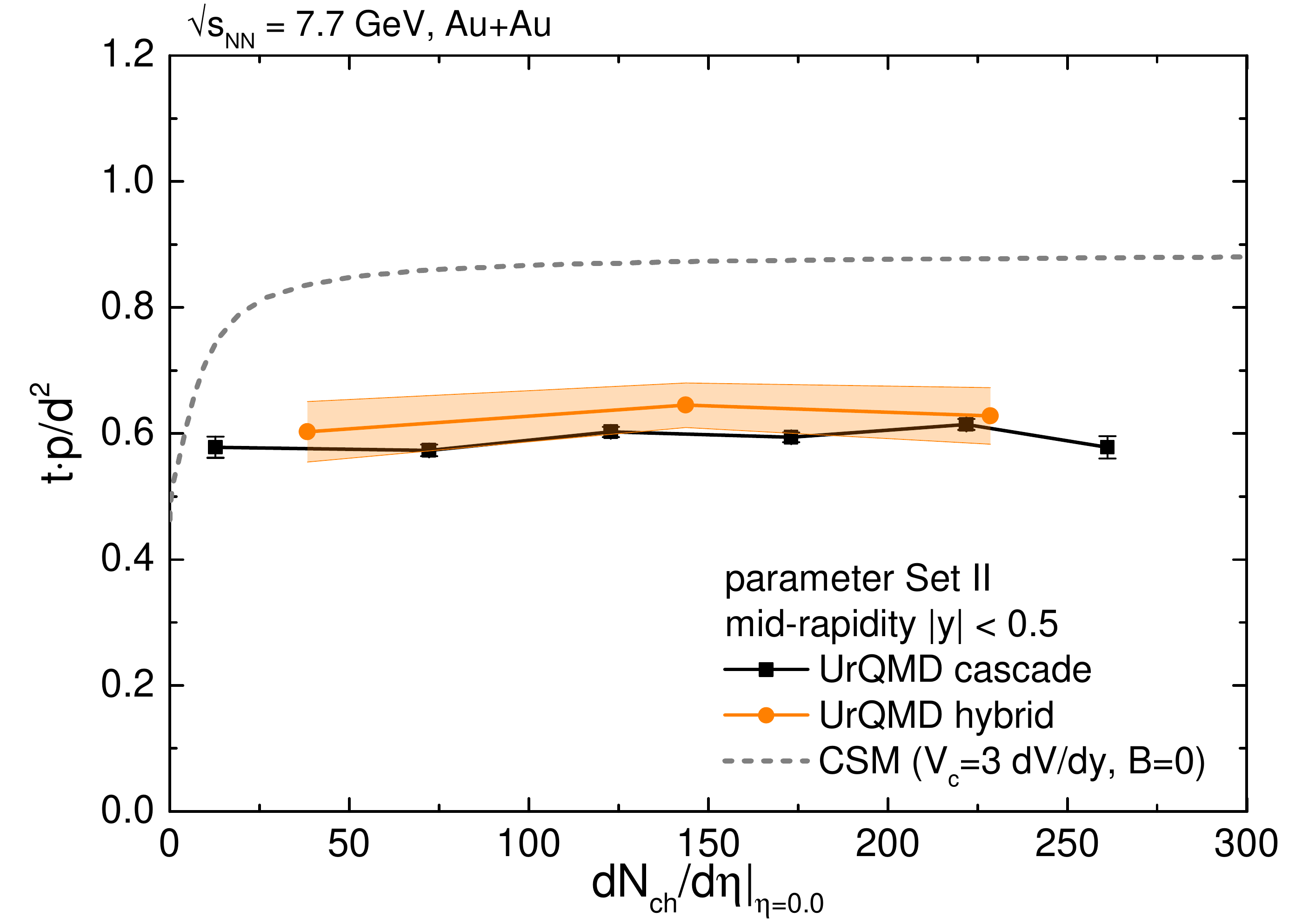}
\caption{[Color online] Centrality dependence of the double ratio tp/d$^2$ for the mid rapidity region of Au+Au collisions at $\sqrt{s_{\mathrm{NN}}}=7.7$ GeV. The ratio for both the UrQMD cascade simulations (black squares) as well as the hybrid model simulations (orange circles) show only a very mild centrality dependence.
The gray dashed line corresponds to the predictions of the canonical statistical model with local baryon conservation at LHC energies.
}\label{cent}
\end{figure}

Let us now investigate the possible breaking of the scaling for small systems. Here, we extend this discussion to peripheral collisions, i.e. smaller system sizes. As has already been observed in the canonical thermal model for ALICE data \cite{Vovchenko:2018fiy}, a decrease of the $d/p^2$ ratio for very peripheral collisions can be expected due to the effects of (local) baryon conservation. 
To explore, if this effect is unique to the canonical thermal description, we explore the volume scaling for peripheral collisions in the coalescence approach. 
In figure \ref{cent_sc} we show the two ratios $d/p^2$ and $t/p^3$, multiplied with the corresponding powers of the charged particle multiplicity as function of centrality for Au+Au collisions at $\sqrt{s_{NN}}=7.7$ GeV. 
Here, it can be clearly seen that the volume scaling only holds for central collisions, while in more peripheral collisions the scaling is broken. 
One may attribute a couple of mechanisms to this effect:
\begin{enumerate}
\item canonical suppression of cluster states or 
\item a change in the chemical and kinetic freeze-out temperature and chemical potential when going to peripheral reactions as well as fluctuations there-off.
\item effects of the small/finite size of the system created in peripheral collisions which leads to less well defined freeze out dynamics.
\end{enumerate}
As the mean chemical freeze-out temperature is rather constant with centrality, strong variations of the temperature are unlikely to be a simple explanation.
The canonical suppression is more promising, although a rather localized range of baryon conservation is required for the effect to be visible even in small systems.
More complex explanations may also involve different inhomogeneous distributions of the temperature and chemical potentials at various centralities~(e.g. due to different collision geometries), which are outside the scope of the present study. 

The system-size effects seem to be equally strong for both, $d/p^2$ and $t/p^3$, ratios in our UrQMD calculations, and once the double ratio $tp/d^2$ is considered, they almost entirely cancel. 
This is shown in figure \ref{cent}, where both the UrQMD cascade model as well as the hydro hybrid model show essentially no change of the double ratio for peripheral collisions. To put this finding into context with the thermal model, the results for a canonical statistical model (CSM), for a centrality independent temperature, is also shown as dashed grey line. The CSM results are obtained for a canonical volume which 3 times larger than the mid rapidity volume and the total baryon number in $V_c$ is set to zero, which corresponds to a system similar to that at LHC energies \footnote{The proper treatment of canonical corrections for low beam energies at high baryon density is significantly more difficult and thus we show the results for LHC energies.}. This CSM result is consistent with canonical thermal model results of \cite{Vovchenko:2018fiy} where only slight decrease of the double ratio for very small systems is observed, which is not seen for the coalescence approach. Note, that this may be due to the fact that at low beam energies the net baryon number, even for very peripheral collisions can be still finite and thus the canonical effects are even weaker here.

\section{Summary}

We have extended the phase-space coalescence approach within UrQMD in its cascade and/or hydro-hybrid version to describe the production of tritons and $^3$He in heavy-ion collisions.
The model parameters were first fixed to data from Au-Au collisions at $E_{\rm lab} = 10.8A$~GeV, and the model was used to study the energy and system-size dependence of light nuclei production for a broad energy range from SIS to LHC.
The phase-space coalescence approach shows qualitative agreement with the experimental results of different experimental collaborations as well with thermal model predictions. 
Sizable deviations from the data is observed at very low beam energies and attributed to
missing fragmentation processes, suggesting that multi-fragmentation plays a crucial role to understand cluster production in that regime. 

The tension in triton production between the STAR and ALICE experiments cannot be explained with the UrQMD or UrQMD+hydro model or the thermal model.
A volume scaling is confirmed for the ratios $\mathrm{d}/\mathrm{p}^2$ and $\mathrm{t}/\mathrm{p}^3$ and is only broken for peripheral collisions in which (micro-) canonical and finite-size effects grow in importance. This also means that results on nuclei scaling properties at the same multiplicity but different beam energy should not necessarily be comparable. 
Importantly, our explicit microscopic treatment of coalescence from UrQMD and hydro-hybrid simulations shows essentially no volume dependence for the double ratio of $\mathrm{t}\mathrm{p}/\mathrm{d}^2$. For beam energies $\mathrm{E_{lab}}< 10 A $ GeV, a rapid increase of the double ratio is observed due to anti-correlation between deuteron and triton multiplicity caused by global conservation and the microscopic treatment of cluster formation. 
This is in contrast to findings of \cite{Zhao:2021dka}, where a clear centrality dependence of the ratio was found, but in agreement with the centrality dependence observed in ALICE data \cite{Vovchenko:2018fiy} and needs to be understood.

\section*{Acknowledgments}
JS thanks the Samson AG and the BMBF through the ErUM-Data project for funding. 
VV acknowledges the support through the
Feodor Lynen Program of the Alexander von Humboldt
foundation, the U.S. Department of Energy, 
Office of Science, Office of Nuclear Physics, under contract number 
DE-AC02-05CH11231231, and within the framework of the
Beam Energy Scan Theory (BEST) Topical Collaboration. 
This work was supported by the DAAD through a PPP exchange grant.
Computational resources were provided by the Frankfurt Center for Scientific Computing (Goethe-HLR). PH acknowledges support by the Frankfurt Institute of Advanced Studies (FIAS).

\end{document}